\documentclass[preprint,12pt]{elsarticle}

\usepackage{graphicx}
\usepackage{txfonts}
\usepackage{dcolumn}
\usepackage{bm}
\usepackage{hyperref}
\usepackage{color}
\usepackage{epsfig}
\usepackage[normalem]{ulem}
\usepackage{amssymb}
\usepackage{natbib}
\biboptions{sort&compress}

\newcommand{\beq}{\begin{equation}}
\newcommand{\eeq}{\end{equation}}
\newcommand{\begg}{\begin{gather}}
\newcommand{\eegg}{\end{gather}}

\journal{Astroparticle Physics}

\begin{document}

\begin{frontmatter}

\title{Solar wind test of the de Broglie-Proca massive photon with Cluster multi-spacecraft data}

   \author{Alessandro Retin\`o\textsuperscript{1}}
\ead{alessandro.retino@lpp.polytechnique.fr}
\ead[url]{http://www.lpp.fr/-Alessandro-Retino-}
\author{Alessandro D.A.M. Spallicci\textsuperscript{2}\corref{ca}}
\ead{spallicci@cnrs-orleans.fr}
\ead[url]{http://lpc2e.cnrs-orleans.fr/$\sim$spallicci/}
\cortext[ca]{Corresponding author}
\author{Andris Vaivads\textsuperscript{3}}
\ead{andris.vaivads@irfu.se}
\ead[url]{http://www.uu.se/sokresultatsida/?q=vaivads}

\address{\textsuperscript{1}Centre National de la Recherche Scientifique, LPP UMR 7648\\
\mbox{Ecole Polytechnique - Universit\'e Pierre et Marie Curie Paris VI - Observatoire de Paris}\\
Route de Saclay 91128 Palaiseau, France\\
\textsuperscript{2}Universit\'e d'Orl\'eans, Observatoire des Sciences de l'Univers en r\'egion Centre, UMS 3116\\
Centre Nationale de la Recherche Scientifique, LPC2E UMR 7328\\
\mbox {3A Av. de la Recherche Scientifique, 45071 Orl\'eans, France}
\\
\textsuperscript{3}Uppsala Universitet, Institutet f\"or Rymdfysik, {\AA}ngstr\"omlaboratoriet\\ 
L\"agerhyddsv\"agen 1, L{\aa}da 537, 751 21 Uppsala, Sverige
}
         
   \date{Received 14 February 2016}

\begin{abstract}
Our understanding of the universe at large and small scales relies largely on electromagnetic observations. As photons are the messengers, fundamental physics has a concern in testing their properties, including the absence of mass. We use Cluster four spacecraft data in the solar wind at 1 AU to estimate the mass upper limit for the photon. We look for deviations from 
Amp\`ere's law, through the curlometer technique for the computation of the magnetic field, and through the measurements of ion and electron velocities for the computation of the current. We show that the upper bound for 
$m_\gamma$ lies between $1.4 \times 10^{-49}$ and $3.4 \times 10^{-51}$ kg, and thereby discuss the currently accepted lower limits in the solar wind.
\end{abstract}


\end{frontmatter}

\section{Introduction}

We have just witnessed the opening of the gravitational wave window on the universe \cite{abbottetal2016,spalliccietal2005}. Nevertheless, our understanding of the universe at large and small scales will continue to rely also on electromagnetic observations. As photons are the messengers, fundamental physics has a concern in testing their properties, including the absence of mass.

Furthermore, while alternative theories of to general relativity are conceived for solving the questions raised by the dark universe or to couple gravity with the other interactions, less effort is deployed for studying alternative electromagnetism. Nevertheless, electromagnetism  may differ from the Maxwellian conception of the nineteenth century and thereby contribute to solve some of the riddles in contemporary physics and cosmology.  

Non-Maxwellian theories can be classed into non-linear or massive photon based. The former were initiated by 
Born and Infeld \cite{boin34}, Heisenberg and Euler \cite{heeu36}. For the latter, the initiator was 
de Broglie who assessed the photon mass lower than $10^{-53}$ kg \cite{db22,db23}, reinterpreted the work of his student Proca \cite{pr36a,pr36b,pr36c,pr36d,pr37}, and achieved a comprehensive formulation of the photon \cite{db40}. 
The de Broglie-Proca, henceforth dBP, equations follow relativity laws for reference frames moving at constant velocities (Lorentz-Poincar\'e transformations); instead, a change of potential implies a change in the field (not Lorenz gauge invariant). 
Gauge invariant formulation have been later proposed by several authors; among the early contributions, we have Stueckelberg \cite{st38a,st38b,st57}, Podolsky \cite
{podolski1942,podolskikikuchi1944,podolskischwed1948}, Chern and Simons \cite{chernsimons1974}. Phenomenologically oriented reviews include \cite{goni10,tulugi05,acciolynetoscatena2010a,spqigiro11}.

Massive photons have been evoked for dark matter, inflation, charge conservation, magnetic monopoles, Higgs boson, redshifts; in applied physics, superconductors and "light shining through walls" experiments. The mass can be considered effective, if depending on given parameters.

How much the foundations of physics are affected by a massive photon is not straightforward to assess, for the variety of the theories, the removal of our ordinary landmarks and the rising of interwoven implications.  

The dBP equations \cite{db40} differ from the four original Maxwell ones 
solely in the divergence of the electric field $\vec {E}$, and in the curl of the magnetic field $\vec {B}$. {They describe a massive spin-1 boson.} In SI units, the dBP equations are 

\beq
{\vec {\nabla}} \cdot \vec {E} = \frac{\rho}{\epsilon_0} - {\cal M}^2\phi~,
\label{gauss1mod}
\eeq\
\beq
{\vec {\nabla}} \times \vec{E} = - \frac{\partial \vec {B}}{\partial t}~,
\label{faraday}
\eeq
\beq
{\vec {\nabla}} \cdot \vec {B} = 0~,
\label{gauss2}
\eeq
\beq
{\vec {\nabla}} \times \vec{B} = \mu_0 \vec{j} + \mu_0 \epsilon_0 \frac{\partial \vec {E}}{\partial t} - {\cal M}^2 \vec{A}~,
\label{amperemaxwellmod}
\eeq
where ${\cal M} = m_\gamma c/{\hbar} = 1/ \lambdabar$, 
$m_\gamma$ is the photon mass, 
$c = 2.99\times 10^8$ m~s$^{-1}$ the speed of light, 
$\hbar = 1.05 \times 10^{-34}$ J s the reduced Planck's constant, 
$\lambdabar$ the reduced Compton wavelength, 
$\epsilon_0 = 8.85 \times 10^{-12}$ F m$^{-1}$ the permittivity, 
$\mu_0 = 1.26 \times 10^{-6}$ N A$^{-2}$ the permeability, 
$\rho$ the charge density, 
$\vec{j}$ the current density vector, 
$\phi$ and $\vec{A}$ the scalar and vector potential.

The first estimates are due to Schr\"odinger \cite{sc43,basc55}, who noted that $\vec{A}$
falls off exponentially at large distances

\beq
{\vec A} = {\vec {\nabla}} \times \frac {\vec m}{r} e^{-r/\lambdabar}~.
\eeq

For a distance comparable to the reduced Compton wavelength, the
exponential term lead to the observable deviations from the Maxwell power law
scaling, equivalent to determining the photon mass.

A laboratory Coulomb's law test determined the mass upper limit of $2\times 10^{-50}$~kg \cite{wifahi71}. 
de Broglie \cite{db40} observed that photon speed would be proportional to the inverse of the square of the frequency, unfortunately 
like {it occurs} for dispersion {due to the ionised parts of the interstellar medium}. For pulsars, delays of lower energy incoming photons are routinely measured, but lacking any 
independent estimate on the electron density, the differences are solely attributed to plasma. The dispersion-based limit is $3\times 10^{-49}$ kg \cite{bawh72}. In the last years, photon dispersion laws have been extensively analysed in the quest for quantum gravity, e.g.,  
\cite{amelino-cameliaetal1998,abdoetal2009,amelino-camelia2009,mavromatos2010,ellismavromatos2013}, 
where classical relativistic symmetries are broken by Planck-scale effects,
with the emergence of a preferred frame, or else \cite{amelinocamelia2002,kowalskiglikman-nowak2002,magueijo-smolin2003,amelinocamelia2013} where relativistic symmetries are only deformed by the Planck scale. 

With Jupiter and Earth magnetic fields, a limit around $10^{-51}$ kg was set \cite{dagoni75,fikllalupe94}. In the solar wind, Ryutov found $10^{-52}$~kg at 1 AU \cite{ry97,ry07}, and $1.5\times 10^{-54}$~kg at $40$ AU \cite{ry07}, 
limit accepted by the Particle Data Group (PDG) \cite{oliveetal2014}. For recent studies of various nature, see 
\cite{acciolynetoscatena2010a,pacogubeis12,qian2012,coma14,dono14,boelmasasgsp2016}. Lower limits ($3 \times 10^{-63}$ kg) have been claimed when modelling the galactic magnetic field \cite{ya59,ch76,addvgr07}. 
The lowest value for any mass is dictated by Heisenberg's  
principle $m \geq \hbar/\Delta t c^2$, and gives $3.8\times 10^{-69}$~kg,
where $\Delta t$ is the supposed age of the Universe.

The large-scale \cite{ya59,ch76,addvgr07} and other limits might be legitimate, but remain the outcomes of models and observations rather than strictly experimental limits.
Indeed, Goldhaber and Nieto state {"Quoted photon-mass limits have at times been overly optimistic in the strengths of their characterizations. This is perhaps due to the temptation to assert too
strongly something one "knows" to be true" \cite{goni10}. We share this concern.  

Estimates from solar wind  magnetic fields \cite{ry97,ry07,liushao2012} are partly based on {\it in situ} measurements, but a close scrutiny reveals that: (i) the magnetic field is assumed exactly always and everywhere a Parker spiral; (ii) the accuracy of particle data measurements  (from, {\it e.g.}, Pioneer or Voyager) has not been discussed; (iii) there is no error analysis, nor data presentation. In \cite{ry97}, Ryutov first refers to the limit of $10^{-51}$ kg, relative to the Jupiter magnetic field, obtained by others in \cite{dagoni75}. Then he discusses that an imbalance of the magnetic forces in the neighbouring areas of the solar wind would have caused violent plasma motions with an average energy exceeding the energy of the ions by three orders of magnitude; since such motions are not observed in the solar wind, the author lowers the estimate in \cite{dagoni75} of one order of magnitude, and sets the mass upper limit at 1 AU. In \cite{ry07}, still Ryutov refers to Voyager 1 and Voyager 2  data appeared in previous work \cite{burlaganesswangsheeley1998,nessburlaga2001,burlaganesswangsheeley2003} to justify strict use of Parker's model, and to adopt the {\it reductio ad absurdum} approach already used in \cite{ry97}. Indeed, on the basis that the magnetic field is almost entirely azimuthal, Ryutov considers that the Lorentz force would be increased of a factor $(L_B/\lambdabar)^2$, where $L_B$ is the magnetic field characteristic length, with respect to the Maxwellian case. Since the deviations from the observed flow
structure would become grossly incompatible with the real situation, the mass upper limit is lowered to $1.5 \times 10^{-54}$ kg at 40 AU. 
A margin of a factor three constitutes the  error budget. Finally, the authors in \cite{liushao2012}, again without presenting data, argue that the mass upper limit could be lowered of a factor two.  

For checking such solar wind estimates, we therefore attempt a more experimental approach. 
We need either a precise experiment or a  
large apparatus, since a small $m_\gamma$ is associated to a  
large $\lambdabar$.   
Herein, we focus on the second possibility through the largest-scale magnetic field accessible to {\it in situ} spacecraft measurements, {\it i.e.} the interplanetary magnetic field carried by the solar wind.  

\begin{figure}
\centering \includegraphics[width=12cm]{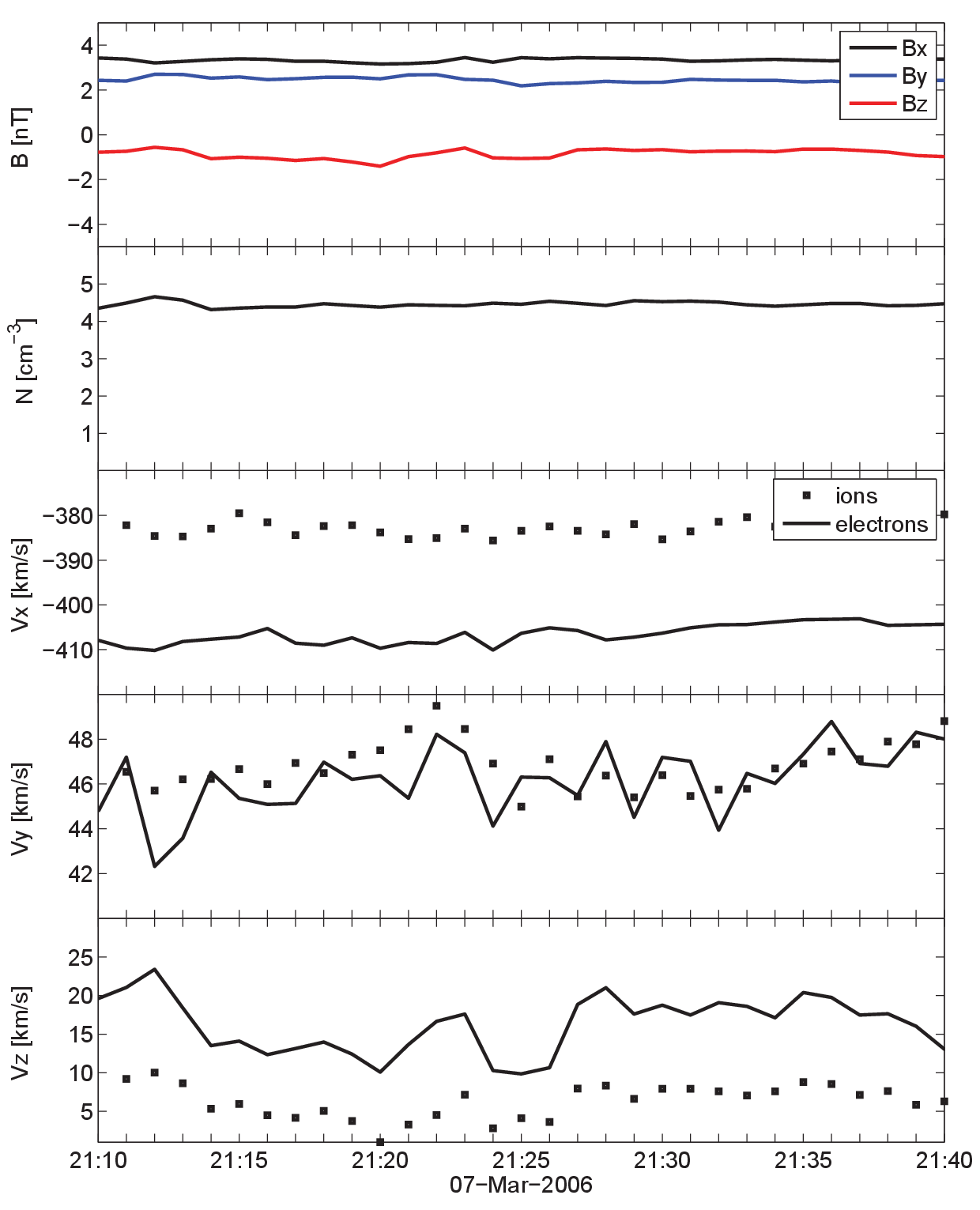}
\caption{Panel ({First from above}). The three components of the magnetic field for Cluster 3 in the GSE (Geocentric Solar Ecliptic) coordinate system. The magnetic field (nT) is measured at 22 samples/s and has been re-sampled at the time resolution of the spacecraft position measurements (1 minute). The GSE coordinate system is defined as having the X direction pointing from the Earth to the Sun, the Z direction orthogonal to the ecliptic plane, and the Y direction completing the right-handed system. 
Panel ({Second from above}). The average plasma density (N cm$^{-3}$). Panels ({Third, Fourth, Fifth from above}). The $v_x, v_x, v_x$ velocity components (
km s$^{-1}$) of ions (dotted line) and electrons (full line) in GSE, respectively. The ion velocity is measured by one spacecraft (C1). The electron velocity is the average over the tetrahedron of the velocity measured by all four spacecraft (C1,C2,C3,C4).}
\label{fig1}
\end{figure}

\section{Dealing with Cluster data}

Cluster \cite{esscgo97} is an ESA mission composed by four spacecraft flying in tetrahedral configuration at $1$ AU from the Sun, and separated by distances ranging from $10^5$ to $10^7$ m. It has allowed the first direct computation of three-dimensional quantities such as ${\vec {\nabla}} \times {\vec B}$ from magnetic field four-point measurements in the solar wind.  

For large-scale steady components of the magnetic field, {\it i.e.} very low frequencies, the displacement current density in Eq. (\ref{amperemaxwellmod}) can be dropped. For $v_{sw}$ the solar wind velocity (for the numerical values, see the text below), we have  
\[
\epsilon_0\mu_0 \frac{\partial E}{\partial t}\sim
\epsilon_0\mu_0 \frac{E v_{sw}}{L_B}\sim 
\epsilon_0\mu_0 \frac{B v_{sw}^2}{L_B}\sim
2 \times 10^{-22}~{\rm~A~m}^{-2}~,
\]
and Eq. (\ref{amperemaxwellmod}) reduces to the dBP Amp\`ere's law

\beq
{\vec {\nabla}} \times \vec{B} = \mu_0 \vec{j} - {\cal M}^2 \vec{A}~.
\label{amperemod}
\eeq

For the current densities $\vec {j}_B = {\vec {\nabla}} \times \vec{B} / \mu_{0} $ and $\vec {j}= \vec{j}_P = n e (\vec{v}_i - \vec{v}_e)$, $n$ the number density {$n = 4.46 \times 10^6$ m$^{-3}$}, 
$e$ the electron charge {$1.6 \times 10^{-19}$ C}, ${\vec v}_i$, ${\vec v}_e$ the velocity of the ions and electrons, respectively, the dBP photon mass is
\vspace{-5pt}
\beq
m_\gamma  =  \frac{k}{\left|\vec{A}_H\right|^{~\frac{1}{2}}}\left|n e (\vec{v}_i - \vec{v}_e) - \frac{{\vec {\nabla}}\times\vec{B}}{\mu_0}\right| ^{~\frac{1}{2}}  = \frac{k \left|\vec{j}_P - \vec{j}_B \right| ^{~\frac{1}{2}}}{\left|\vec{A}_H\right|^{~\frac{1}{2}}}~,
\label{idealmeas}
\eeq
where $k = \hbar \mu_0^{~\frac{1}{2}}c^{-1}$, and $\vec{A}_H$ is the vector potential from the interplanetary magnetic field.

We have selected a few events in Cluster solar wind orbits
with the following criteria: (i) an {undisturbed} solar wind, {\it i.e.} disconnected from the terrestrial bow shock, and as far as possible from the terrestrial magnetic field; (ii) the closest location of the spacecraft to the equatorial plane; (iii) the widest inter-spacecraft separation, about $10^7$~m,  
assuring the largest differences in the magnetic field among the spacecraft; (iv) the configuration best approaching the tetrahedron; (v) the availability of good quality particle currents.
We discuss the event having the most accurate particle data (7 March 2006) since the particle current density is the observable that eventually determines the upper bound. Figure  \ref{fig1} shows the magnetic field measured by Cluster for this event. 
The magnetic field has $B_x>0$, $B_y>0$ and $B_x, B_y \gg B_z$, as expected for a Parker spiral configuration close to the ecliptic plane. It is emphasised, however, that our analysis does not rely on the Parker model, since the magnetic field is measured {\it in situ}. The choice of an event having Parker's spiral orientation is to have conditions as similar as possible to those presented in \cite{ry97,ry07} which are the only available estimates on the upper bound of $m_\gamma$ in the solar wind.

We measure $j_B$ using the curlometer technique \cite{dubaglro02breve,roduroch98breve} on the magnetic field data from the fluxgate magnetometer \cite{baetal97breve}. The average ${\vec {\nabla}} \times \vec{B} $ is computed over the tetrahedron with no assumptions on the field analytical form (only assuming linear gradients). The error on $j_B$ is %
\vspace{-3pt}
\beq
\frac{\Delta j_B}{<j_B>} = \frac{\Delta B}{<\delta B>} + \frac{\Delta R}{<R>}~,
\label{errcurl}
\eeq
with $<\!j_B\!>$ being the average current density, $\Delta B$ the error on the magnetic field, 
$<\!\delta B\!>$ the average value of the magnetic field difference between the satellites, $\Delta {R}$ the error on the spacecraft separation $R$, $<\!R\!>$ the average separation between the satellites.
An estimate on the deviation from linearity is given by the quantity
$|{\vec {\nabla}} \cdot {\vec B}| /|{\vec {\nabla}} \times {\vec B}| $.
The magnitudes of $j_{B}$ and of its absolute error $\Delta j_{B}$ are shown in Fig. \ref{fig4}. The
absolute errors on the magnetic field $\Delta {B} = 0.1$ nT and on spacecraft separation $\Delta {R} = 10^3$ m correspond to typical values \cite{roduroch98breve}.
The average current density over the duration of the event is $<\!j_B\!>{\approx}~(3.5 \pm 4.7)\times 10^{-11}$ A~m$^{-2}$.  
The relative error and the value of $|{\vec {\nabla}} \cdot {\vec B}| / | {\vec {\nabla}} \times {\vec B}| {\approx} 
75\%$ are larger than typical estimates \cite{roduroch98breve}. This is due to the large values of the elongation $ E \sim 0.6$ and the planarity $ P\sim 0.7$ of the tetrahedron for this event (the ideal estimate occurs for a perfect tetrahedron with E=P=0). In absence of an event with both a correct tetrahedron shape and adequate particle measurements, we picked an event where the latter condition 
is met (conversely, we found a much better accuracy for $<\!j_B\!>$ in other events - not shown herein).  Similar independent results on the error on  $<\!j_B\!>$, through applying random variations on magnetic field measurement and satellite position at each of the satellites, and estimating the standard deviation of the obtained current fluctuations.

%

The particle current density is $<\!j_P\!> {\approx }~(1.86 \pm 3) \times 10^{-8}$ A~m$^{-2}$, Fig. \ref{fig4}. In all events considered, we have not found significantly lower values of $j_{P}$ and of its error. An accurate assessment is difficult due to inherent instrument limitations. Ion electrostatic analysers \cite{reetal01} can saturate in the solar wind due to the presence of high fluxes, resulting often in incorrect ion currents. The computation of the currents from electron analyzers is affected by photoelectrons and spacecraft charging issues  \cite{joetal97}. The value of the current density $j_P$ for this event (as well as for other events) is much larger than that from the curlometer. This is mostly due to the differences in the velocities, while the estimate on density is reasonable. Figure \ref{fig4} shows two (for other events up to four) orders of magnitude between $j_P$ and $j_B$. While we can't exclude that (part of) this difference is {due to the dBP massive photon}, the large uncertainties of particle currents hint forcefully to instrumental limits. 

\subsection{The estimations of $A_H$}

Since Eq. (\ref{gauss2}) h, holds, then   
${\vec {\nabla}} \times {\vec B} = {\vec A}$ still holds in dBP theory in absence of magnetic monopoles. Nevertheless, Eq. (\ref{amperemod}) depends on the value of ${\vec A}$. Thus, conversely to Maxwellian electromagnetism, the dBP theory is not {Lorenz} gauge invariant. The potential is {perceived as} a measurable quantity, the values of which implies a change in the fields. Turning to $A_{H}$ due to the interplanetary magnetic field at Cluster location, we arrange three estimations.   

\subsubsection{Four point estimate}

From $j_{B}$ {and $B = 4.23 \times 10^{-9}$ T}, we get $L_{B} \sim B / \mu_{0} j_{B} \approx 9.6 \times 10^7$ m. For our event, the inter-spacecraft separation is $ 6\times 10^6$ m, thus much smaller than $L_{B}$. The characteristic value of the vector potential is $A_{H} \sim B \times L_{B} \approx 4.1 \times 10^{-1}$ T m. The variation $\delta$  of this  estimate is comparable to the observables themselves $B \sim \delta A_{H} / \delta L \sim A_{H} / L_{B}$ \cite{ry07}.

The errors on $B$ and $j_B$ lead to even lower values of $A_H$ {($1.1 \times 10^{-1}$ T m)}, and thus a higher upper limit on $m_\gamma$ {of a factor two}. 
%

The advantages of the curlometer method consist in performing instantaneous measurements, in avoiding assumptions on the local structure of $B$ and on the (or lack of) steadiness of the solar wind. On the other hand, it is based on the first derivatives of $B$, and thus it 
{may not be }suited for small volumes.

\subsubsection{Single spacecraft estimate}

Alternatively $A_{H}$ is found {through} a single spacecraft {monitoring} $B$ as it advects past the spacecraft. The drawback is that $A_{H}$ can only be evaluated along the solar wind flow (approximately, the GSE $X$ direction), as opposed to the curlometer estimate being performed in all directions. A different aspects of the same limit, the single spacecraft method works exclusively in presence of temporal variations in the solar wind. Further, without an analytical model of the magnetic field and thus of $A_H$, from only {\it in situ} measurements it is not possible to determine the direction where the largest component of $A_H$ lies. 
We assessed the Z component of $A_{H}$ due to the Y component of $B$ (being $B^Y > B^Z$) over an interval of about $\pm$ 12 hours around the event pointed in Fig. \ref {fig1}. 
{For the event close to the equatorial plane 
\beq
B^Y = \frac{\partial A_H^Z}{\partial x} - \frac{\partial A_H^X}{\partial z} \sim \frac{\partial A_H^Z}{\partial x}~,
\eeq
we get}
\beq
\delta A^Z_H \sim A^Z_H \sim\int B^Y(t) dx \sim v_{sw}\int B^Y (t) dt~,  
\label{integration}
\eeq
being $v_{sw}= 4 \times
10^5$ m s$^{-1}$ the typical advection velocity. It appears that $A_{H}$ is linearly changing in the time interval 14:00-22:00, where $B{^Y}$ is roughly constant, 
and thus
$A^Z_H \approx {29}$ T m. Along the  flow direction, it determines $L_B \approx 6.7 \times 10^{9} $ m, about seventy times larger than $L_B$ obtained through the curlometer.  

It is not possible to establish which of these two approaches is more reliable. 

\subsubsection{Parker's model estimate}

We wish to investigate in all directions, and we thus recur to the Parker model for a third (non-experimental, {\it i.e.} without data) assessment. {We compute $A_H$ in spherical coordinates from \cite{bieberetal1987}
\beq
A_r  = \frac{2b}{3}\left[ 1 - \frac{3}{2} x - x \ln (1 + x)\right]~, 
\label{bieberr}
\eeq 
\beq
A_\theta = \frac{2b}{3}\sin \theta\left[ \frac{x}{1+x}+ \ln (1 + x)\right] \left ( \frac{\cos \theta}{x}\right )~,
\label{biebertheta}
\eeq
\beq
A_\phi = \frac{a}{r \sin \theta}(1 - x)~, 
\label{bieberphi}
\eeq
where $ x = |\cos \theta|$, $a = 3.54 \times 10^{-9}$ T AU$^2$ and $b = 3.54 \times 10^{-9}$ T AU, } and get $A_H =637$ T m (for $\theta = \pi/2$).

{The vector potential, Eqs. (\ref{bieberr}-\ref{bieberphi}), is computed in Coulomb's gauge $\nabla\cdot {\vec A} = 0$. The dBP equations  are not Lorenz gauge invariant but automatically satisfy the Lorenz gauge, that is $\nabla\cdot {\vec A} + 1/c^2 \partial \phi/\partial t = 0$. Thus, in our Coulomb gauge case, the scalar potential $\phi$ must be constant in time. This latter condition inserted in the time derivative of ${\vec E} = \nabla{\phi } - \partial {\vec A}/\partial t$, and recalling that we deal with a static case for which $\partial {\vec E}/\partial t=0$, implies that Eqs. (\ref{bieberr}-\ref{bieberphi}) are valid only if $A_H$ varies at most linearly in time. Indeed, the event under scrutiny fulfills this feature.}  

\begin{figure}
\centering \includegraphics[width=14cm]{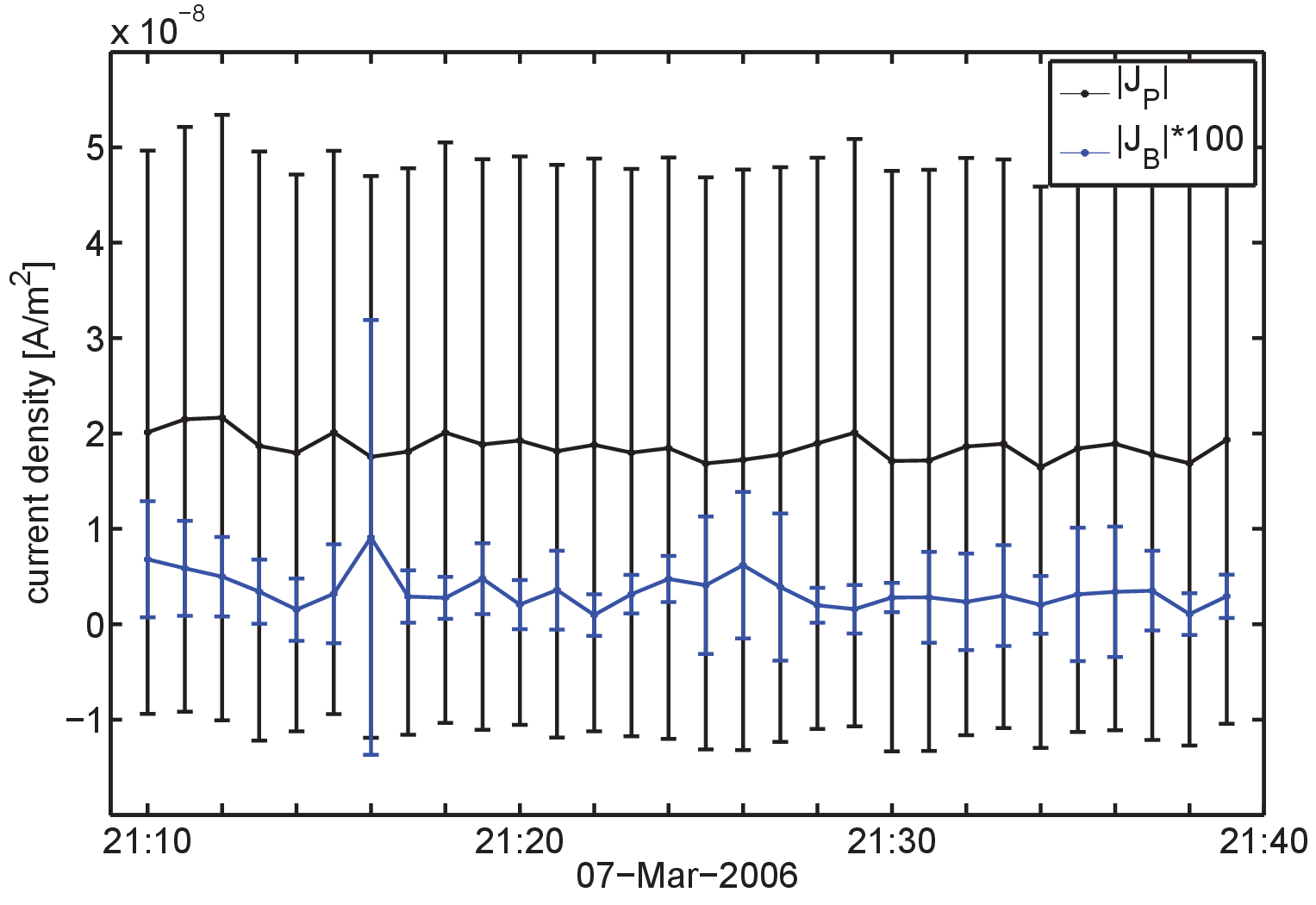}
\caption{The current densities $j_{B}$ (computed with the curlometer) and $j_{P}$ (inferred from the particle detectors) and the associated experimental errors. $\Delta {j}_{B}$ is estimated by using Eq. (\ref{errcurl}), while $\Delta j_{P}$ {is} obtained {through} the propagation of maximal errors on density n, ion velocity $v_{i}$ and electron velocity $v_{e}$ taken as
$\Delta {n} = 10\% $,
$\Delta v_{i} {=} \Delta v_{e} = 5\%$ \cite{pafasc98}.}
\label{fig4}
\end{figure}

\subsection{The mass upper limits}
 
We retain from the previous sections that the value of $A_H$ is result of estimates and not of a direct measurement. At best, it depends indirectly on measurements (four point and single spacecraft estimates); else, it is the outcome of a model (Parker estimate). 
Further, the values of $A_H$ itself vary greatly in the three estimates. Thus, considering its errors would appear misplaced.   
Indeed, it seems appropriate that our experimental approach measures an upper limit for ${A_H}^{\frac{1}{2}} m_\gamma$, Eq. (\ref{idealmeas}), and that we compute 
three estimates of $m_\gamma$, one for each value of $A_H$.
Finally, we consider $k$ an error-free constant (in dBP theory $c$ depends on the frequency, but herein we don't deal with dispersion issues). 

Error propagation analysis indicates for the worst case \cite{taylor1997}      

\[
{A_H}^{\frac{1}{2}} \left (m_\gamma + \Delta m_\gamma \right)  =  {A_H}^{\frac{1}{2}} \left (m_\gamma + \left| \frac {\partial m_\gamma}{\partial j_P}\right | \Delta j_P + \left| \frac {\partial m_\gamma}{\partial j_B}\right | \Delta j_B\right ) = 
\]
\beq
 ~k \left[(j_P - j_B)^{\frac{1}{2}}+ \frac{\Delta j_P + \Delta j_B}{2(j_P - j_B)^{\frac{1}{2}}}\right]~. 
\label{masslimit1}
\eeq




Considering $j_{P}$ and $\Delta j_{P}$ of the same order, $j_P = 0.62~\Delta j_P$, and both much larger than $j_{B}$ and $\Delta j_{B}$, Eq. (\ref{masslimit1}), after squaring, obviously leads to 

\beq
{A_H}^{\frac{1}{2}} \left (m_\gamma + \Delta m_\gamma \right) \sim  k \left(j_{P} + \Delta j_{P}\right)^{1/2} ~.
\label{masslimit3}
\eeq

The outcome, Eq. (\ref{masslimit3}), can be drawn directly by Eq. (\ref{idealmeas}) using the same previous assumptions.
Table (\ref{tab1}) displays the values of $m_\gamma$, for three different estimated values of $A_H$. The upper limits are all well above those appeared in \cite{ry97,ry07}. 

\begin{table}
\centering
\caption{The values of $m_\gamma$ (according to the estimate on $A_H$). }
\label{tab1}
\begin{tabular}{@{}|l|c|c|c|}
\hline
$A_H$ [T m] & 0.4 & {29} (Z) & 637 \\ \hline
$m_\gamma$ [kg] & $1.4 \times 10^{-49}$ 
                & {$1.6 \times 10^{-50}$}
                & $3.4 \times 10^{-51}$ \\
\hline
\end{tabular}
\end{table}

\section{Discussion}

The most stringent limitation comes from the particle detectors. The difference between ion and electron velocities is
$v_{i-e} \sim (j_P + \Delta j_P)/n e = 6.8 \times 10^4$ ms$^{-1}$.
Recasting Eq. (\ref{masslimit3}), we 
derive $v_{i-e}$ that particle detectors should measure to resolve the $m_\gamma$ upper bound for a given $A_H$
\vspace{-3pt}
\beq
v_{i-e} \sim \frac{A_H m_\gamma^2}{k^2 n e}~.
\label{vie}
\eeq
This is plotted in Fig. \ref{fig5as} where the vertical lines refer to earlier literature and to our Cluster spacecraft test.  
The upper limit $10^{-52}$ kg in \cite{ry97,ry07} requires resolving a difference $v_{i-e}$ of approximately $3.6\times 10^{-2}$, ${2.6}$, $5.7 \times 10^{1}$  m s$^{-1}$ (for the three values of $A_H$) which is not possible with currently available particle detectors, onboard Cluster or other spacecraft. 
Otherwise stated, one would need $A_H = 7.5 \times 10^5$ T m, to obtain an upper limit of $10^{-52}$ kg with the measured $v_{i-e} \approx 6.8 \times 10^{4}$ m s$^{-1}$. Such high values of $A_H$ are not stated in \cite{ry97,ry07};  conversely a value of $10^3$ T m is stated in \cite{ry09}. 

{We further note that the upper limit in \cite{ry97,ry07} of $10^{-52}$ kg requires a very low value of $j_P + \Delta j_P$, even for the highest value of the potential, that is $A_H = 637$ T m. Thus, assuming the Parker model and $j_P + \Delta j_P \simeq j_B + \Delta_B \sim  10^{-11}$  A~m$^{-2}$, we get the currently accepted limits at 1 AU. Nevertheless, we don't find such low values of particle current in our data.}  

Few improvements could still be achieved with Cluster data. One could select only those components of $j_P$ having the best accuracy. Figure \ref{fig1} shows the differences between ion and electron velocities in the GSE coordinate system. 
The smallest difference $\approx 10^3$ m s$^{-1}$ is in the Y direction, {\it i.e.}, in the equatorial plane. 
This is the most accurate component for this event, also considering that both ion and electron velocities are close to the solar wind velocity along  Y  due to solar wind aberration (on average $\approx 3 \times 10^4$ m s$^{-1}$). When using $j^Y_{P}= (4.21 \pm 33.97) \times 10^{-10}$ A m$^{-2}$, we lower the $m_\gamma$ upper limit of a factor four{, and $v_{i-e}^Y$ is around $5.3 \times 10^{3}$ m s$^{-1}$}. However, this implies that $A_{H}$ has its largest component in the Y direction, as indicated by the Parker model.  

\begin{figure}
\centering \includegraphics[width=14cm]{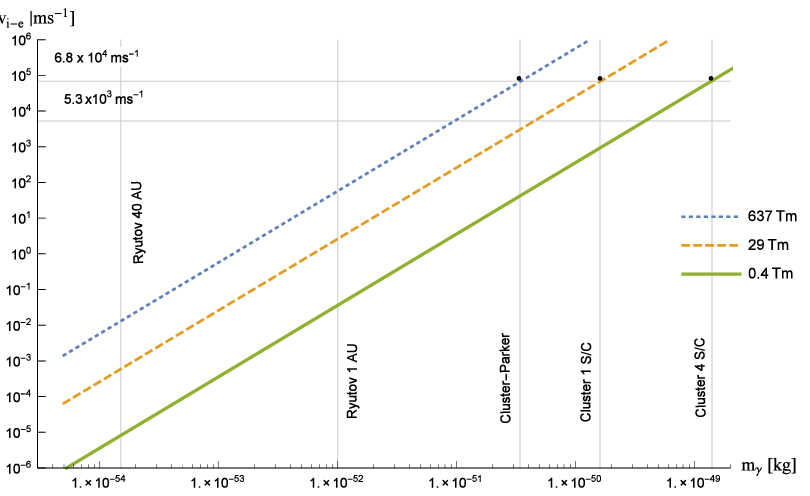}
\caption{The ion-electron velocity difference $v_{i-e}$ versus $m_\gamma$, for three $A_H$ values ($0.4, {29}, 637$ Tm). The vertical dotted lines represent solar wind 
\cite{ry97,ry07} and Cluster limits, Tab. (\ref{tab1}). The horizontal dotted {lines represent $v_{i-e}=6.8 \times 10^4$ and $v_{i-e}^Y = 5.3 \times 10^3$ m s$^{-1}$} for Cluster.}
\label{fig5as}
\end{figure}

Our analysis does not assume that the solar wind is in steady-state (Parker's model), except for the third estimate on $A_H$. Therefore, our approach is indeed different from \cite{ry97,ry07}. Therein, the average large scale properties of the solar wind are taken from statistical datasets; and $m_\gamma$ is obtained through {\it reductio ad absurdum} arguments on the integral effect of magnetic stresses (${\vec j} \times {\vec B}$) with respect to the dynamic solar wind pressure, see Eq. (11) in \cite{ry07}, assuming a steady-state Parker's model of the magnetic field. 
In our approach, we estimate the local vector potential $A_H$, and we measure the local currents $j_B$, $j_P$ with few assumptions on the structure of the magnetic field and on the steadiness of the solar wind. This implies that our estimates on $A_H$, notably the former two, are not only related to the large scale average field described by Parker's model, but also to the always present fluctuations in the solar wind.

\section{Conclusions}

We have reported three new estimates for the upper limit of the dBP photon mass by using Cluster multi-spacecraft measurements in the solar wind. We have found larger 
values than previous estimates, our test being based on fewer assumptions.  
 First, we directly assess ${\vec {\nabla}} \times \vec{B}/\mu_0 $ from four-point magnetic field measurements; this was before impossible with a single spacecraft.  
Second, for two estimates we  
do not assume that the interplanetary magnetic field is a Parker spiral, though we have chosen events compatible to the Parker spiral for comparison with earlier findings \cite{ry97,ry07}. For the third estimate, we assume the Parker model, and yet we get a larger upper limit than in \cite{ry97,ry07}.
 Third, it is the only solar wind test that takes into account in detail the experimental errors. The domain between our findings ($m_\gamma < 10^{-49}$ kg) and the results from {\it ad hoc} model in the solar wind ($m_\gamma < 10^{-52}, 10^{-54}$ kg) is still subjected to assumptions and conjectures, though fewer now, and not to clear-cutting outcomes from experiments.
Our experiment is limited by the resolution of the velocity difference between ions and electrons. 

\section{Perspectives}

For achieving more reliable results, 
the improvement of particle detector performances and planning multi-spacecraft mission appear both mandatory.
Already with Cluster, a resolution of $10^3$ m s$^{-1}$ is attainable when considering the best measured Y component of the flow that nicely reproduces the expected aberration of the solar wind. Further improvements could come from better estimates on the moments in the solar wind, that is mainly electrons which are more affected by the spacecraft potential. 

The MMS 
satellites might represent another opportunity to apply our method, given the features of the mission and of the measured currents \cite{grahametal2015}. The impact of the limited separation distance can be assessed, and a study of the impact of the time and velocity resolutions of the particle detectors on the upper limit of the photon mass would be desirable.  

We hope that payload designers will be attentive to the requirement that fundamental physics measurements pose. If so, new avenues of research could be explored. 

\section{Acknowledgements}
We thank E. Adelberger, C. Carr, A. Chasapis, A. Fazakerley, P. Henri, J.-P. Lebreton, V. Krasnoselskikh, F. Piazza, J.-L. Pin\c{c}on, P. Robert,  and the Cluster Active Archive. ADAMS as  French Chair at the Departamento de F\'isica Te{\'o}rica of the Universidade do Estado do Rio de Janeiro thanks UERJ and the French {Consulate}. 

\bibliographystyle{model1-num-names.bst} 
\bibliography{references_spallicci_160406}

\begin{thebibliography}{64}
\expandafter\ifx\csname natexlab\endcsname\relax\def\natexlab#1{#1}\fi
\providecommand{\url}[1]{\texttt{#1}}
\providecommand{\href}[2]{#2}
\providecommand{\path}[1]{#1}
\providecommand{\DOIprefix}{doi:}
\providecommand{\ArXivprefix}{arXiv:}
\providecommand{\URLprefix}{URL: }
\providecommand{\Pubmedprefix}{pmid:}
\providecommand{\doi}[1]{\href{http://dx.doi.org/#1}{\path{#1}}}
\providecommand{\Pubmed}[1]{\href{pmid:#1}{\path{#1}}}
\providecommand{\bibinfo}[2]{#2}
\ifx\xfnm\relax \def\xfnm[#1]{\unskip,\space#1}\fi
\bibitem[{Abbott et~al.(2016)Abbott, Abbott, Abbott, Abernathy, Acernese,
  Ackley, Adams, Adams, Addesso, Adhikari, Adya, Affeldt, Agathos, Agatsuma,
  Aggarwal, Aguiar, Aiello, Ain, Ajith, Allen, Allocca, Altin, Anderson,
  Anderson, Arai, Arain, Araya, Arceneaux, Areeda, Arnaud, Arun, Ascenzi,
  Ashton, Ast, Aston, Astone, Aufmuth, Aulbert, Babak, Bacon, Bader, Baker,
  Baldaccini, Ballardin, Ballmer, Barayoga, Barclay, Barish, Barker, Barone,
  Barr, Barsotti, Barsuglia, Barta, Bartlett, Barton, Bartos, Bassiri, Basti,
  Batch, Baune, Bavigadda, Bazzan, Behnke, Bejger, Belczynski, Bell, Bell,
  Berger, Bergman, Bergmann, Berry, Bersanetti, Bertolini, Betzwieser, Bhagwat,
  Bhandare, Bilenko, Billingsley, Birch, Birney, Birnholtz, Biscans, Bisht,
  Bitossi, Biwer, Bizouard, Blackburn, Blair, Blair, Blair, Bloemen, Bock,
  Bodiya, Boer, Bogaert, Bogan, Bohe, Bojtos, Bond, Bondu, Bonnand, Boom, Bork,
  Boschi, Bose, Bouffanais, Bozzi, Bradaschia, Brady, Braginsky, Branchesi,
  Brau, Briant, Brillet, Brinkmann, Brisson, Brockill, Brooks, Brown, Brown,
  Brown, Buchanan, Buikema, Bulik, Bulten, Buonanno, Buskulic, Buy, Byer,
  Cabero, Cadonati, Cagnoli, Cahillane, Bustillo, Callister, Calloni, Camp,
  Cannon, Cao, Capano, Capocasa, Carbognani, Caride, Diaz, Casentini, Caudill,
  Cavagli\`a, Cavalier, Cavalieri, Cella, Cepeda, {Cerboni Baiardi}, Cerretani,
  Cesarini, Chakraborty, Chalermsongsak, Chamberlin, Chan, Chao, Charlton,
  Chassande-Mottin, and et~al.}]{abbottetal2016}
\bibinfo{author}{B.~P. Abbott}, \bibinfo{author}{R.~Abbott},
  \bibinfo{author}{T.~D. Abbott}, \bibinfo{author}{M.~R. Abernathy},
  \bibinfo{author}{F.~Acernese}, \bibinfo{author}{K.~Ackley},
  \bibinfo{author}{C.~Adams}, \bibinfo{author}{T.~Adams},
  \bibinfo{author}{P.~Addesso}, \bibinfo{author}{R.~X. Adhikari},
  \bibinfo{author}{V.~B. Adya}, \bibinfo{author}{C.~Affeldt},
  \bibinfo{author}{M.~Agathos}, \bibinfo{author}{K.~Agatsuma},
  \bibinfo{author}{N.~Aggarwal}, \bibinfo{author}{O.~D. Aguiar},
  \bibinfo{author}{L.~Aiello}, \bibinfo{author}{A.~Ain},
  \bibinfo{author}{P.~Ajith}, \bibinfo{author}{B.~Allen},
  \bibinfo{author}{A.~Allocca}, \bibinfo{author}{P.~A. Altin},
  \bibinfo{author}{S.~B. Anderson}, \bibinfo{author}{W.~G. Anderson},
  \bibinfo{author}{K.~Arai}, \bibinfo{author}{M.~A. Arain},
  \bibinfo{author}{M.~C. Araya}, \bibinfo{author}{C.~C. Arceneaux},
  \bibinfo{author}{J.~S. Areeda}, \bibinfo{author}{N.~Arnaud},
  \bibinfo{author}{K.~G. Arun}, \bibinfo{author}{S.~Ascenzi},
  \bibinfo{author}{G.~Ashton}, \bibinfo{author}{M.~Ast}, \bibinfo{author}{S.~M.
  Aston}, \bibinfo{author}{P.~Astone}, \bibinfo{author}{P.~Aufmuth},
  \bibinfo{author}{C.~Aulbert}, \bibinfo{author}{S.~Babak},
  \bibinfo{author}{P.~Bacon}, \bibinfo{author}{M.~K.~M. Bader},
  \bibinfo{author}{P.~T. Baker}, \bibinfo{author}{F.~Baldaccini},
  \bibinfo{author}{G.~Ballardin}, \bibinfo{author}{S.~W. Ballmer},
  \bibinfo{author}{J.~C. Barayoga}, \bibinfo{author}{S.~E. Barclay},
  \bibinfo{author}{B.~C. Barish}, \bibinfo{author}{D.~Barker},
  \bibinfo{author}{F.~Barone}, \bibinfo{author}{B.~Barr},
  \bibinfo{author}{L.~Barsotti}, \bibinfo{author}{M.~Barsuglia},
  \bibinfo{author}{D.~Barta}, \bibinfo{author}{J.~Bartlett},
  \bibinfo{author}{M.~A. Barton}, \bibinfo{author}{I.~Bartos},
  \bibinfo{author}{R.~Bassiri}, \bibinfo{author}{A.~Basti},
  \bibinfo{author}{J.~C. Batch}, \bibinfo{author}{C.~Baune},
  \bibinfo{author}{V.~Bavigadda}, \bibinfo{author}{M.~Bazzan},
  \bibinfo{author}{B.~Behnke}, \bibinfo{author}{M.~Bejger},
  \bibinfo{author}{C.~Belczynski}, \bibinfo{author}{A.~S. Bell},
  \bibinfo{author}{C.~J. Bell}, \bibinfo{author}{B.~K. Berger},
  \bibinfo{author}{J.~Bergman}, \bibinfo{author}{G.~Bergmann},
  \bibinfo{author}{C.~P.~L. Berry}, \bibinfo{author}{D.~Bersanetti},
  \bibinfo{author}{A.~Bertolini}, \bibinfo{author}{J.~Betzwieser},
  \bibinfo{author}{S.~Bhagwat}, \bibinfo{author}{R.~Bhandare},
  \bibinfo{author}{I.~A. Bilenko}, \bibinfo{author}{G.~Billingsley},
  \bibinfo{author}{J.~Birch}, \bibinfo{author}{R.~Birney},
  \bibinfo{author}{O.~Birnholtz}, \bibinfo{author}{S.~Biscans},
  \bibinfo{author}{A.~Bisht}, \bibinfo{author}{M.~Bitossi},
  \bibinfo{author}{C.~Biwer}, \bibinfo{author}{M.~A. Bizouard},
  \bibinfo{author}{J.~K. Blackburn}, \bibinfo{author}{C.~D. Blair},
  \bibinfo{author}{D.~G. Blair}, \bibinfo{author}{R.~M. Blair},
  \bibinfo{author}{S.~Bloemen}, \bibinfo{author}{O.~Bock},
  \bibinfo{author}{T.~P. Bodiya}, \bibinfo{author}{M.~Boer},
  \bibinfo{author}{G.~Bogaert}, \bibinfo{author}{C.~Bogan},
  \bibinfo{author}{A.~Bohe}, \bibinfo{author}{P.~Bojtos},
  \bibinfo{author}{C.~Bond}, \bibinfo{author}{F.~Bondu},
  \bibinfo{author}{R.~Bonnand}, \bibinfo{author}{B.~A. Boom},
  \bibinfo{author}{R.~Bork}, \bibinfo{author}{V.~Boschi},
  \bibinfo{author}{S.~Bose}, \bibinfo{author}{Y.~Bouffanais},
  \bibinfo{author}{A.~Bozzi}, \bibinfo{author}{C.~Bradaschia},
  \bibinfo{author}{P.~R. Brady}, \bibinfo{author}{V.~B. Braginsky},
  \bibinfo{author}{M.~Branchesi}, \bibinfo{author}{J.~E. Brau},
  \bibinfo{author}{T.~Briant}, \bibinfo{author}{A.~Brillet},
  \bibinfo{author}{M.~Brinkmann}, \bibinfo{author}{V.~Brisson},
  \bibinfo{author}{P.~Brockill}, \bibinfo{author}{A.~F. Brooks},
  \bibinfo{author}{D.~A. Brown}, \bibinfo{author}{D.~D. Brown},
  \bibinfo{author}{N.~M. Brown}, \bibinfo{author}{C.~C. Buchanan},
  \bibinfo{author}{A.~Buikema}, \bibinfo{author}{T.~Bulik},
  \bibinfo{author}{H.~J. Bulten}, \bibinfo{author}{A.~Buonanno},
  \bibinfo{author}{D.~Buskulic}, \bibinfo{author}{C.~Buy},
  \bibinfo{author}{R.~L. Byer}, \bibinfo{author}{M.~Cabero},
  \bibinfo{author}{L.~Cadonati}, \bibinfo{author}{G.~Cagnoli},
  \bibinfo{author}{C.~Cahillane}, \bibinfo{author}{J.~C. Bustillo},
  \bibinfo{author}{T.~Callister}, \bibinfo{author}{E.~Calloni},
  \bibinfo{author}{J.~B. Camp}, \bibinfo{author}{K.~C. Cannon},
  \bibinfo{author}{J.~Cao}, \bibinfo{author}{C.~D. Capano},
  \bibinfo{author}{E.~Capocasa}, \bibinfo{author}{F.~Carbognani},
  \bibinfo{author}{S.~Caride}, \bibinfo{author}{J.~C. Diaz},
  \bibinfo{author}{C.~Casentini}, \bibinfo{author}{S.~Caudill},
  \bibinfo{author}{M.~Cavagli\`a}, \bibinfo{author}{F.~Cavalier},
  \bibinfo{author}{R.~Cavalieri}, \bibinfo{author}{G.~Cella},
  \bibinfo{author}{C.~B. Cepeda}, \bibinfo{author}{L.~{Cerboni Baiardi}},
  \bibinfo{author}{G.~Cerretani}, \bibinfo{author}{E.~Cesarini},
  \bibinfo{author}{R.~Chakraborty}, \bibinfo{author}{T.~Chalermsongsak},
  \bibinfo{author}{S.~J. Chamberlin}, \bibinfo{author}{M.~Chan},
  \bibinfo{author}{S.~Chao}, \bibinfo{author}{P.~Charlton},
  \bibinfo{author}{E.~Chassande-Mottin}, \bibinfo{author}{et~al.},
\newblock \bibinfo{title}{Observation of gravitational waves from a binary
  black hole merger},
\newblock \bibinfo{journal}{Phys. Rev. Lett.} \bibinfo{volume}{116}
  (\bibinfo{year}{2016}) \bibinfo{pages}{061102}.
\bibitem[{Spallicci et~al.(2005)Spallicci, Aoudia, {de Freitas Pacheco},
  Regimbau, and Frossati}]{spalliccietal2005}
\bibinfo{author}{A.~Spallicci}, \bibinfo{author}{S.~Aoudia},
  \bibinfo{author}{J.~{de Freitas Pacheco}}, \bibinfo{author}{T.~Regimbau},
  \bibinfo{author}{G.~Frossati},
\newblock \bibinfo{title}{Virgo detector optimization for gravitational waves
  by inspiralling binaries},
\newblock \bibinfo{journal}{Class. Q. Grav.} \bibinfo{volume}{22}
  (\bibinfo{year}{2005}) \bibinfo{pages}{S461}.
\bibitem[{Born and Infeld(1934)}]{boin34}
\bibinfo{author}{M.~Born}, \bibinfo{author}{L.~Infeld},
\newblock \bibinfo{title}{Foundations of the new field theory},
\newblock \bibinfo{journal}{Proc. R. Soc. Lond. A} \bibinfo{volume}{144}
  (\bibinfo{year}{1934}) \bibinfo{pages}{425}.
\bibitem[{Heisenberg and Euler(1936)}]{heeu36}
\bibinfo{author}{W.~Heisenberg}, \bibinfo{author}{H.~Euler},
\newblock \bibinfo{title}{Folgerungen aus der {D}iracschen theorie des
  positron},
\newblock \bibinfo{journal}{Z. Phys.} \bibinfo{volume}{98}
  (\bibinfo{year}{1936}) \bibinfo{pages}{714}.
\bibitem[{{de Broglie}(1922)}]{db22}
\bibinfo{author}{L.~{de Broglie}},
\newblock \bibinfo{title}{Rayonnement noir et quanta de lumi\`ere},
\newblock \bibinfo{journal}{J. Phys. et Radium} \bibinfo{volume}{VI 3}
  (\bibinfo{year}{1922}) \bibinfo{pages}{422}.
\bibitem[{{de Broglie}(1923)}]{db23}
\bibinfo{author}{L.~{de Broglie}},
\newblock \bibinfo{title}{Ondes et quanta},
\newblock \bibinfo{journal}{Comptes Rendus Hebd. S{\'e}ances Acad. Sc. Paris}
  \bibinfo{volume}{177} (\bibinfo{year}{1923}) \bibinfo{pages}{507}.
\bibitem[{Proca(1936{\natexlab{a}})}]{pr36a}
\bibinfo{author}{A.~Proca},
\newblock \bibinfo{title}{Sur la th\'eorie du positon},
\newblock \bibinfo{journal}{Comptes Rendus Hebd. S{\'e}ances Acad. Sc. Paris}
  \bibinfo{volume}{202} (\bibinfo{year}{1936}{\natexlab{a}})
  \bibinfo{pages}{1366}.
\bibitem[{Proca(1936{\natexlab{b}})}]{pr36b}
\bibinfo{author}{A.~Proca},
\newblock \bibinfo{title}{Sur les \'equations fondamentales des particules
  \'elementaires},
\newblock \bibinfo{journal}{Comptes Rendus Hebd. S{\'e}ances Acad. Sc. Paris}
  \bibinfo{volume}{202} (\bibinfo{year}{1936}{\natexlab{b}})
  \bibinfo{pages}{1490}.
\bibitem[{Proca(1936{\natexlab{c}})}]{pr36c}
\bibinfo{author}{A.~Proca},
\newblock \bibinfo{title}{Sur les photons et les particules charge pure},
\newblock \bibinfo{journal}{Comptes Rendus Hebd. S{\'e}ances Acad. Sc. Paris}
  \bibinfo{volume}{203} (\bibinfo{year}{1936}{\natexlab{c}})
  \bibinfo{pages}{709}.
\bibitem[{Proca(1936{\natexlab{d}})}]{pr36d}
\bibinfo{author}{A.~Proca},
\newblock \bibinfo{title}{Sur la th\'eorie ondulatoire des \'electrons positifs
  et n\'egatifs},
\newblock \bibinfo{journal}{J. Phys. et Radium} \bibinfo{volume}{VII}
  (\bibinfo{year}{1936}{\natexlab{d}}) \bibinfo{pages}{347}.
\bibitem[{Proca(1937)}]{pr37}
\bibinfo{author}{A.~Proca},
\newblock \bibinfo{title}{Particules libres photons et particules charge pure},
\newblock \bibinfo{journal}{J. Phys. et Radium} \bibinfo{volume}{VIII}
  (\bibinfo{year}{1937}) \bibinfo{pages}{23}.
\bibitem[{{de Broglie}(1940)}]{db40}
\bibinfo{author}{L.~{de Broglie}}, \bibinfo{title}{La m\'echanique ondulatoire
  du photon, Une novelle th\'eorie de la lumi\'ere},
  \bibinfo{publisher}{Hermann}, \bibinfo{address}{Paris}, \bibinfo{year}{1940}.
\bibitem[{Stueckelberg(1938{\natexlab{a}})}]{st38a}
\bibinfo{author}{E.~C.~G. Stueckelberg},
\newblock \bibinfo{title}{Die wechselwirkungskr{\"a}fte in der elektrodynamik
  und in der feldtheorie der kernkr{\"a}fte ({T}eil {I})},
\newblock \bibinfo{journal}{Helv. Phys. Acta} \bibinfo{volume}{11}
  (\bibinfo{year}{1938}{\natexlab{a}}) \bibinfo{pages}{225}.
\bibitem[{Stueckelberg(1938{\natexlab{b}})}]{st38b}
\bibinfo{author}{E.~C.~G. Stueckelberg},
\newblock \bibinfo{title}{Die wechselwirkungskr{\"a}fte in der elektrodynamik
  und in der feldtheorie der kernkr{\"a}fte ({T}eil {II} und {III})},
\newblock \bibinfo{journal}{Helv. Phys. Acta} \bibinfo{volume}{11}
  (\bibinfo{year}{1938}{\natexlab{b}}) \bibinfo{pages}{299}.
\bibitem[{Stueckelberg(1957)}]{st57}
\bibinfo{author}{E.~C.~G. Stueckelberg},
\newblock \bibinfo{title}{Th\'eorie de la radiation de photons de masse
  arbitrairement petite},
\newblock \bibinfo{journal}{Helv. Phys. Acta} \bibinfo{volume}{30}
  (\bibinfo{year}{1957}) \bibinfo{pages}{209}.
\bibitem[{Podolsky(1942)}]{podolski1942}
\bibinfo{author}{B.~Podolsky},
\newblock \bibinfo{title}{A generalized electrodynamics. {P}art {I} -
  non-quantum},
\newblock \bibinfo{journal}{Phys. Rev.} \bibinfo{volume}{62}
  (\bibinfo{year}{1942}) \bibinfo{pages}{68}.
\bibitem[{Podolsky and Kikuchi(1944)}]{podolskikikuchi1944}
\bibinfo{author}{B.~Podolsky}, \bibinfo{author}{C.~Kikuchi},
\newblock \bibinfo{title}{A generalized electrodynamics. {P}art {II} -
  quantum},
\newblock \bibinfo{journal}{Phys. Rev.} \bibinfo{volume}{65}
  (\bibinfo{year}{1944}) \bibinfo{pages}{228}.
\bibitem[{Podolsky and Schwed(1948)}]{podolskischwed1948}
\bibinfo{author}{B.~Podolsky}, \bibinfo{author}{P.~Schwed},
\newblock \bibinfo{title}{Review of a generalized electrodynamics},
\newblock \bibinfo{journal}{Rev. Mod. Phys.} \bibinfo{volume}{20}
  (\bibinfo{year}{1948}) \bibinfo{pages}{40}.
\bibitem[{Chern and Simons(1974)}]{chernsimons1974}
\bibinfo{author}{S.~Chern}, \bibinfo{author}{J.~Simons},
\newblock \bibinfo{title}{Characteristic forms and geometric invariants},
\newblock \bibinfo{journal}{Ann. Math.} \bibinfo{volume}{99}
  (\bibinfo{year}{1974}) \bibinfo{pages}{48}.
\bibitem[{Goldhaber and Nieto(2010)}]{goni10}
\bibinfo{author}{A.~S. Goldhaber}, \bibinfo{author}{M.~M. Nieto},
\newblock \bibinfo{title}{Photon and graviton mass limits},
\newblock \bibinfo{journal}{Rev. Mod. Phys.} \bibinfo{volume}{82}
  (\bibinfo{year}{2010}) \bibinfo{pages}{939}.
\bibitem[{{Tu} et~al.(2005){Tu}, {Luo}, and {Gillies}}]{tulugi05}
\bibinfo{author}{L.-C. {Tu}}, \bibinfo{author}{J.~{Luo}},
  \bibinfo{author}{G.~T. {Gillies}},
\newblock \bibinfo{title}{The mass of the photon},
\newblock \bibinfo{journal}{Rep. Progr. Phys.} \bibinfo{volume}{68}
  (\bibinfo{year}{2005}) \bibinfo{pages}{77}.
\bibitem[{Accioly et~al.(2010)Accioly, Helay{\"e}l-Neto, and
  Scatena}]{acciolynetoscatena2010a}
\bibinfo{author}{A.~Accioly}, \bibinfo{author}{J.~Helay{\"e}l-Neto},
  \bibinfo{author}{E.~Scatena},
\newblock \bibinfo{title}{Upper bounds on the photon mass},
\newblock \bibinfo{journal}{Phys. Rev. D} \bibinfo{volume}{82}
  (\bibinfo{year}{2010}) \bibinfo{pages}{065026}.
\bibitem[{Spavieri et~al.(2011)Spavieri, Quintero, Gillies, and
  Rodriguez}]{spqigiro11}
\bibinfo{author}{G.~Spavieri}, \bibinfo{author}{J.~Quintero},
  \bibinfo{author}{G.~T. Gillies}, \bibinfo{author}{M.~Rodriguez},
\newblock \bibinfo{title}{A survey of existing and proposed classical and
  quantum approaches to the photon mass},
\newblock \bibinfo{journal}{Eur. Phys. J. D} \bibinfo{volume}{61}
  (\bibinfo{year}{2011}) \bibinfo{pages}{531}.
\bibitem[{Schr{\"o}dinger(1943)}]{sc43}
\bibinfo{author}{E.~Schr{\"o}dinger},
\newblock \bibinfo{title}{The {E}arth's and the {S}un's permanent magnetic
  fields in the unitary field theory},
\newblock \bibinfo{journal}{Proc. Royal Irish Acad. A} \bibinfo{volume}{49}
  (\bibinfo{year}{1943}) \bibinfo{pages}{135}.
\bibitem[{Bass and Schr{\"o}dinger(1955)}]{basc55}
\bibinfo{author}{L.~Bass}, \bibinfo{author}{E.~Schr{\"o}dinger},
\newblock \bibinfo{title}{Must the photon mass be zero?},
\newblock \bibinfo{journal}{Proc. R. Soc. London A Math. Phys. Sc.}
  \bibinfo{volume}{232} (\bibinfo{year}{1955}) \bibinfo{pages}{1}.
\bibitem[{Williams et~al.(1971)Williams, Faller, and Hill}]{wifahi71}
\bibinfo{author}{E.~R. Williams}, \bibinfo{author}{J.~E. Faller},
  \bibinfo{author}{H.~A. Hill},
\newblock \bibinfo{title}{New experimental test of {C}oulomb's law: A
  laboratory upper limit on the photon rest mass},
\newblock \bibinfo{journal}{Phys. Rev. Lett.} \bibinfo{volume}{26}
  (\bibinfo{year}{1971}) \bibinfo{pages}{721}.
\bibitem[{Bay and White(1972)}]{bawh72}
\bibinfo{author}{Z.~Bay}, \bibinfo{author}{J.~A. White},
\newblock \bibinfo{title}{Frequency dependence of the speed of light in space},
\newblock \bibinfo{journal}{Phys. Rev. D} \bibinfo{volume}{5}
  (\bibinfo{year}{1972}) \bibinfo{pages}{796}.
\bibitem[{{Amelino{-}Camelia} et~al.(1998){Amelino{-}Camelia}, Ellis,
  Mavromatos, Nanopoulos, and Sarkar}]{amelino-cameliaetal1998}
\bibinfo{author}{G.~{Amelino{-}Camelia}}, \bibinfo{author}{J.~Ellis},
  \bibinfo{author}{N.~E. Mavromatos}, \bibinfo{author}{D.~V. Nanopoulos},
  \bibinfo{author}{S.~Sarkar},
\newblock \bibinfo{title}{Tests of quantum gravity from observations of
  {$\gamma$}-ray bursts},
\newblock \bibinfo{journal}{Nat.} \bibinfo{volume}{393} (\bibinfo{year}{1998})
  \bibinfo{pages}{763}. \bibinfo{note}{{C}orrection, 1998, Nat., 395, 525}.
\bibitem[{Abdo et~al.(2009)Abdo, Ackermann, Ajello, Asano, Atwood, Axelsson,
  Baldini, Ballet, Barbiellini, Baring, Bastieri, Bechtol, Bellazzini, Berenji,
  Bhat, Bissaldi, Bloom, Bonamente, Bonnell, Borgland, Bouvier, Bregeon, Brez,
  Briggs, Brigida, Bruel, Burgess, Burnett, Caliandro, Cameron, Caraveo,
  Casandjian, Cecchi, Çelik, Chaplin, Charles, Cheung, Chiang, Ciprini, Claus,
  Cohen-Tanugi, Cominsky, Connaughton, Conrad, Cutini, Dermer, de~Angelis,
  de~Palma, Digel, Dingus, do~Couto~e Silva, Drell, Dubois, Dumora, Farnier,
  Favuzzi, Fegan, Finke, Fishman, Focke, Foschini, Fukazawa, Funk, Fusco,
  Gargano, Gasparrini, Gehrels, Germani, Gibby, Giebels, Giglietto, Giordano,
  Glanzman, Godfrey, Granot, Greiner, Grenier, Grondin, Grove, Grupe,
  Guillemot, Guiriec, Hanabata, Harding, Hayashida, Hays, Hoversten, Hughes,
  J{\'o}hannesson, Johnson, Johnson, Johnson, Kamae, Katagiri, Kataoka, Kawai,
  Kerr, Kippen, Kn{\"o}dlseder, Kocevski, Kouveliotou, Kuehn, Kuss, Lande,
  Latronico, Lemoine-Goumard, Longo, Loparco, Lott, Lovellette, Lubrano,
  Madejski, Makeev, Mazziotta, McBreen, McEnery, McGlynn, M{\'e}sz{\'a}ros,
  Meurer, Michelson, Mitthumsiri, Mizuno, Moiseev, Monte, Monzani, Moretti,
  Morselli, Moskalenko, Murgia, Nakamori, Nolan, Norris, Nuss, Ohno, Ohsugi,
  Omodei, Orlando, Ormes, Ozaki, Paciesas, Paneque, Panetta, Parent, Pelassa,
  Pepe, {Pesce-Rollins}, Petrosian, Piron, Porter, Preece, Rain{\`o},
  {Ramirez-Ruiz}, Rando, Razzano, Razzaque, Reimer, Reimer, Reposeur, Ritz,
  Rochester, and et~al.}]{abdoetal2009}
\bibinfo{author}{A.~A. Abdo}, \bibinfo{author}{M.~Ackermann},
  \bibinfo{author}{M.~Ajello}, \bibinfo{author}{K.~Asano},
  \bibinfo{author}{W.~B. Atwood}, \bibinfo{author}{M.~Axelsson},
  \bibinfo{author}{L.~Baldini}, \bibinfo{author}{J.~Ballet},
  \bibinfo{author}{G.~Barbiellini}, \bibinfo{author}{M.~G. Baring},
  \bibinfo{author}{D.~Bastieri}, \bibinfo{author}{K.~Bechtol},
  \bibinfo{author}{R.~Bellazzini}, \bibinfo{author}{B.~Berenji},
  \bibinfo{author}{P.~N. Bhat}, \bibinfo{author}{E.~Bissaldi},
  \bibinfo{author}{E.~D. Bloom}, \bibinfo{author}{E.~Bonamente},
  \bibinfo{author}{J.~Bonnell}, \bibinfo{author}{A.~W. Borgland},
  \bibinfo{author}{A.~Bouvier}, \bibinfo{author}{J.~Bregeon},
  \bibinfo{author}{A.~Brez}, \bibinfo{author}{M.~S. Briggs},
  \bibinfo{author}{M.~Brigida}, \bibinfo{author}{P.~Bruel},
  \bibinfo{author}{J.~M. Burgess}, \bibinfo{author}{T.~H. Burnett},
  \bibinfo{author}{G.~A. Caliandro}, \bibinfo{author}{R.~A. Cameron},
  \bibinfo{author}{P.~A. Caraveo}, \bibinfo{author}{J.~M. Casandjian},
  \bibinfo{author}{C.~Cecchi}, \bibinfo{author}{{\"O}.~Çelik},
  \bibinfo{author}{V.~Chaplin}, \bibinfo{author}{E.~Charles},
  \bibinfo{author}{C.~C. Cheung}, \bibinfo{author}{J.~Chiang},
  \bibinfo{author}{S.~Ciprini}, \bibinfo{author}{R.~Claus},
  \bibinfo{author}{J.~Cohen-Tanugi}, \bibinfo{author}{L.~R. Cominsky},
  \bibinfo{author}{V.~Connaughton}, \bibinfo{author}{J.~Conrad},
  \bibinfo{author}{S.~Cutini}, \bibinfo{author}{C.~D. Dermer},
  \bibinfo{author}{A.~de~Angelis}, \bibinfo{author}{F.~de~Palma},
  \bibinfo{author}{S.~W. Digel}, \bibinfo{author}{B.~L. Dingus},
  \bibinfo{author}{E.~do~Couto~e Silva}, \bibinfo{author}{P.~S. Drell},
  \bibinfo{author}{R.~Dubois}, \bibinfo{author}{D.~Dumora},
  \bibinfo{author}{C.~Farnier}, \bibinfo{author}{C.~Favuzzi},
  \bibinfo{author}{S.~J. Fegan}, \bibinfo{author}{J.~Finke},
  \bibinfo{author}{G.~Fishman}, \bibinfo{author}{W.~B. Focke},
  \bibinfo{author}{L.~Foschini}, \bibinfo{author}{Y.~Fukazawa},
  \bibinfo{author}{S.~Funk}, \bibinfo{author}{P.~Fusco},
  \bibinfo{author}{F.~Gargano}, \bibinfo{author}{D.~Gasparrini},
  \bibinfo{author}{N.~Gehrels}, \bibinfo{author}{S.~Germani},
  \bibinfo{author}{L.~Gibby}, \bibinfo{author}{B.~Giebels},
  \bibinfo{author}{N.~Giglietto}, \bibinfo{author}{F.~Giordano},
  \bibinfo{author}{T.~Glanzman}, \bibinfo{author}{G.~Godfrey},
  \bibinfo{author}{J.~Granot}, \bibinfo{author}{J.~Greiner},
  \bibinfo{author}{I.~A. Grenier}, \bibinfo{author}{M.-H. Grondin},
  \bibinfo{author}{J.~E. Grove}, \bibinfo{author}{D.~Grupe},
  \bibinfo{author}{L.~Guillemot}, \bibinfo{author}{S.~Guiriec},
  \bibinfo{author}{Y.~Hanabata}, \bibinfo{author}{A.~K. Harding},
  \bibinfo{author}{M.~Hayashida}, \bibinfo{author}{E.~Hays},
  \bibinfo{author}{E.~A. Hoversten}, \bibinfo{author}{R.~E. Hughes},
  \bibinfo{author}{G.~J{\'o}hannesson}, \bibinfo{author}{A.~S. Johnson},
  \bibinfo{author}{R.~P. Johnson}, \bibinfo{author}{W.~N. Johnson},
  \bibinfo{author}{T.~Kamae}, \bibinfo{author}{H.~Katagiri},
  \bibinfo{author}{J.~Kataoka}, \bibinfo{author}{N.~Kawai},
  \bibinfo{author}{M.~Kerr}, \bibinfo{author}{R.~M. Kippen},
  \bibinfo{author}{J.~Kn{\"o}dlseder}, \bibinfo{author}{D.~Kocevski},
  \bibinfo{author}{C.~Kouveliotou}, \bibinfo{author}{F.~Kuehn},
  \bibinfo{author}{M.~Kuss}, \bibinfo{author}{J.~Lande},
  \bibinfo{author}{L.~Latronico}, \bibinfo{author}{M.~Lemoine-Goumard},
  \bibinfo{author}{F.~Longo}, \bibinfo{author}{F.~Loparco},
  \bibinfo{author}{B.~Lott}, \bibinfo{author}{M.~N. Lovellette},
  \bibinfo{author}{P.~Lubrano}, \bibinfo{author}{G.~M. Madejski},
  \bibinfo{author}{A.~Makeev}, \bibinfo{author}{M.~N. Mazziotta},
  \bibinfo{author}{S.~McBreen}, \bibinfo{author}{J.~E. McEnery},
  \bibinfo{author}{S.~McGlynn}, \bibinfo{author}{P.~M{\'e}sz{\'a}ros},
  \bibinfo{author}{C.~Meurer}, \bibinfo{author}{P.~F. Michelson},
  \bibinfo{author}{W.~Mitthumsiri}, \bibinfo{author}{T.~Mizuno},
  \bibinfo{author}{A.~A. Moiseev}, \bibinfo{author}{C.~Monte},
  \bibinfo{author}{M.~E. Monzani}, \bibinfo{author}{E.~Moretti},
  \bibinfo{author}{A.~Morselli}, \bibinfo{author}{I.~V. Moskalenko},
  \bibinfo{author}{S.~Murgia}, \bibinfo{author}{T.~Nakamori},
  \bibinfo{author}{P.~L. Nolan}, \bibinfo{author}{J.~P. Norris},
  \bibinfo{author}{E.~Nuss}, \bibinfo{author}{M.~Ohno},
  \bibinfo{author}{T.~Ohsugi}, \bibinfo{author}{N.~Omodei},
  \bibinfo{author}{E.~Orlando}, \bibinfo{author}{J.~F. Ormes},
  \bibinfo{author}{M.~Ozaki}, \bibinfo{author}{W.~S. Paciesas},
  \bibinfo{author}{D.~Paneque}, \bibinfo{author}{J.~H. Panetta},
  \bibinfo{author}{D.~Parent}, \bibinfo{author}{V.~Pelassa},
  \bibinfo{author}{M.~Pepe}, \bibinfo{author}{M.~{Pesce-Rollins}},
  \bibinfo{author}{V.~Petrosian}, \bibinfo{author}{F.~Piron},
  \bibinfo{author}{T.~A. Porter}, \bibinfo{author}{R.~Preece},
  \bibinfo{author}{S.~Rain{\`o}}, \bibinfo{author}{E.~{Ramirez-Ruiz}},
  \bibinfo{author}{R.~Rando}, \bibinfo{author}{M.~Razzano},
  \bibinfo{author}{S.~Razzaque}, \bibinfo{author}{A.~Reimer},
  \bibinfo{author}{O.~Reimer}, \bibinfo{author}{T.~Reposeur},
  \bibinfo{author}{S.~Ritz}, \bibinfo{author}{L.~S. Rochester},
  \bibinfo{author}{et~al.},
\newblock \bibinfo{title}{A limit on the variation of the speed of light
  arising from quantum gravity effects},
\newblock \bibinfo{journal}{Nat.} \bibinfo{volume}{462} (\bibinfo{year}{2009})
  \bibinfo{pages}{331}.
\bibitem[{{Amelino{-}Camelia}(2009)}]{amelino-camelia2009}
\bibinfo{author}{G.~{Amelino{-}Camelia}},
\newblock \bibinfo{title}{Burst of support for relativity},
\newblock \bibinfo{journal}{Nat.} \bibinfo{volume}{462} (\bibinfo{year}{2009})
  \bibinfo{pages}{291}.
\bibitem[{Mavromatos(2010)}]{mavromatos2010}
\bibinfo{author}{N.~E. Mavromatos},
\newblock \bibinfo{title}{String quantum gravity, {L}orentz-invariance
  violation and gamma ray astronomy},
\newblock \bibinfo{journal}{Int. J. Mod. Phys. A} \bibinfo{volume}{25}
  (\bibinfo{year}{2010}) \bibinfo{pages}{5409}.
\bibitem[{Ellis and Mavromatos(2013)}]{ellismavromatos2013}
\bibinfo{author}{J.~Ellis}, \bibinfo{author}{N.~E. Mavromatos},
\newblock \bibinfo{title}{Probes of {L}orentz violation},
\newblock \bibinfo{journal}{Astrop. Physics} \bibinfo{volume}{43}
  (\bibinfo{year}{2013}) \bibinfo{pages}{50}.
\bibitem[{{Amelino-Camelia}(2002)}]{amelinocamelia2002}
\bibinfo{author}{G.~{Amelino-Camelia}},
\newblock \bibinfo{title}{Relativity in space-times with short-distance
  structure governed by an observer-independent ({P}lanckian) length scale},
\newblock \bibinfo{journal}{Int. J. Mod. Phys. D} \bibinfo{volume}{11}
  (\bibinfo{year}{2002}) \bibinfo{pages}{35}.
\bibitem[{{Kowalski-Glikman} and Nowak(2002)}]{kowalskiglikman-nowak2002}
\bibinfo{author}{J.~{Kowalski-Glikman}}, \bibinfo{author}{S.~Nowak},
\newblock \bibinfo{title}{Doubly special relativity theories as different bases
  of {\it {k}}-{P}oincar{\'e} algebra},
\newblock \bibinfo{journal}{Phys. Lett. B} \bibinfo{volume}{539}
  (\bibinfo{year}{2002}) \bibinfo{pages}{126}.
\bibitem[{Magueijo and Smolin(2003)}]{magueijo-smolin2003}
\bibinfo{author}{J.~Magueijo}, \bibinfo{author}{L.~Smolin},
\newblock \bibinfo{title}{Generalized {L}orentz invariance with an invariant
  energy scale},
\newblock \bibinfo{journal}{Phys. Rev. D} \bibinfo{volume}{67}
  (\bibinfo{year}{2003}) \bibinfo{pages}{044017}.
\bibitem[{{Amelino-Camelia}(2013)}]{amelinocamelia2013}
\bibinfo{author}{G.~{Amelino-Camelia}},
\newblock \bibinfo{title}{Quantum spacetime phenomenology},
\newblock \bibinfo{journal}{Liv. Rev. Rel.} \bibinfo{volume}{16}
  (\bibinfo{year}{2013}).
\bibitem[{Davis-jr. et~al.(1975)Davis-jr., Goldhaber, and Nieto}]{dagoni75}
\bibinfo{author}{L.~Davis-jr.}, \bibinfo{author}{A.~S. Goldhaber},
  \bibinfo{author}{M.~N. Nieto},
\newblock \bibinfo{title}{Limit on the photon mass deduced from {P}ioneer-10
  observations of {J}upiter's magnetic field},
\newblock \bibinfo{journal}{Phys. Rev. Lett.} \bibinfo{volume}{35}
  (\bibinfo{year}{1975}) \bibinfo{pages}{1402}.
\bibitem[{Fischbach et~al.(1994)Fischbach, Kloor, Langel, Lui, and
  Peredo}]{fikllalupe94}
\bibinfo{author}{E.~Fischbach}, \bibinfo{author}{H.~Kloor},
  \bibinfo{author}{R.~A. Langel}, \bibinfo{author}{A.~T.~Y. Lui},
  \bibinfo{author}{M.~Peredo},
\newblock \bibinfo{title}{New geomagnetic limits on the photon mass and on
  long-range forces coexisting with electromagnetism},
\newblock \bibinfo{journal}{Phys. Rev. Lett.} \bibinfo{volume}{73}
  (\bibinfo{year}{1994}) \bibinfo{pages}{514}.
\bibitem[{Ryutov(1997)}]{ry97}
\bibinfo{author}{D.~D. Ryutov},
\newblock \bibinfo{title}{The role of finite photon mass in
  magnetohydrodynamics of space plasmas},
\newblock \bibinfo{journal}{Plasma Phys. Contr. Fus.} \bibinfo{volume}{39}
  (\bibinfo{year}{1997}) \bibinfo{pages}{A73}.
\bibitem[{Ryutov(2007)}]{ry07}
\bibinfo{author}{D.~D. Ryutov},
\newblock \bibinfo{title}{Using plasma physics to weigh the photon},
\newblock \bibinfo{journal}{Plasma Phys. Contr. Fus.} \bibinfo{volume}{49}
  (\bibinfo{year}{2007}) \bibinfo{pages}{B429}.
\bibitem[{{Olive} and {the Particle Data Group}(2014)}]{oliveetal2014}
\bibinfo{author}{K.~A. {Olive}}, \bibinfo{author}{{the Particle Data Group}},
\newblock \bibinfo{title}{Review of particle physics},
\newblock \bibinfo{journal}{Chin. Phys. C} \bibinfo{volume}{38}
  (\bibinfo{year}{2014}) \bibinfo{pages}{090001}.
\bibitem[{Pani et~al.(2012)Pani, Cardoso, Gualtieri, Berti, and
  Ishibashi}]{pacogubeis12}
\bibinfo{author}{P.~Pani}, \bibinfo{author}{V.~Cardoso},
  \bibinfo{author}{L.~Gualtieri}, \bibinfo{author}{E.~Berti},
  \bibinfo{author}{A.~Ishibashi},
\newblock \bibinfo{title}{Black-hole bombs and photon-mass bounds},
\newblock \bibinfo{journal}{Phys. Rev. Lett.} \bibinfo{volume}{109}
  (\bibinfo{year}{2012}) \bibinfo{pages}{131102}.
\bibitem[{Qian(2012)}]{qian2012}
\bibinfo{author}{L.~Qian},
\newblock \bibinfo{title}{Constraining photon mass by energy-dependent
  gravitational light bending},
\newblock \bibinfo{journal}{Sc. China Phys. Mech. Astr.} \bibinfo{volume}{55}
  (\bibinfo{year}{2012}) \bibinfo{pages}{523}.
\bibitem[{Colafrancesco and Marchegiani(2014)}]{coma14}
\bibinfo{author}{S.~Colafrancesco}, \bibinfo{author}{P.~Marchegiani},
\newblock \bibinfo{title}{Probing photon decay with the {S}unyaev-{Z}el'dovich
  effect},
\newblock \bibinfo{journal}{Astron. Astrophys.} \bibinfo{volume}{562}
  (\bibinfo{year}{2014}) \bibinfo{pages}{L2}.
\bibitem[{Dolgov and Novikov(2014)}]{dono14}
\bibinfo{author}{A.~D. Dolgov}, \bibinfo{author}{V.~A. Novikov},
\newblock \bibinfo{title}{A cosmological bound on e{\textsuperscript{+}}
  e{\textsuperscript{-}} mass difference},
\newblock \bibinfo{journal}{Phys. Lett. B} \bibinfo{volume}{732}
  (\bibinfo{year}{2014}) \bibinfo{pages}{244}.
\bibitem[{Bonetti et~al.(2016)Bonetti, Ellis, Mavromatos, Sakharov,
  {Sarkisian-Grinbaum}, and Spallicci}]{boelmasasgsp2016}
\bibinfo{author}{L.~Bonetti}, \bibinfo{author}{J.~Ellis},
  \bibinfo{author}{N.~E. Mavromatos}, \bibinfo{author}{A.~S. Sakharov},
  \bibinfo{author}{E.~K. {Sarkisian-Grinbaum}}, \bibinfo{author}{A.~D. A.~M.
  Spallicci}, \bibinfo{title}{Photon mass limits from fast radio bursts},
  \bibinfo{year}{2016}. \href{http://arxiv.org/abs/1602.09135
  [astro-ph.HE]}{\tt arXiv:1602.09135 [astro-ph.HE]}.
\bibitem[{Yamaguchi(1959)}]{ya59}
\bibinfo{author}{Y.~Yamaguchi},
\newblock \bibinfo{title}{A composite theory of elementary particles},
\newblock \bibinfo{journal}{Progr. Theor. Phys. Suppl.} \bibinfo{volume}{11}
  (\bibinfo{year}{1959}) \bibinfo{pages}{1}.
\bibitem[{Chibisov(1976)}]{ch76}
\bibinfo{author}{G.~V. Chibisov},
\newblock \bibinfo{title}{Astrophysical upper limits on the photon rest mass},
\newblock \bibinfo{journal}{Sov. Phys. Usp.} \bibinfo{volume}{19}
  (\bibinfo{year}{1976}) \bibinfo{pages}{624}. \bibinfo{note}{[{U}sp. Fiz.
  Nauk, 119 (1976) 591]}.
\bibitem[{Adelberger et~al.(2007)Adelberger, Dvali, and Gruzinov}]{addvgr07}
\bibinfo{author}{E.~Adelberger}, \bibinfo{author}{G.~Dvali},
  \bibinfo{author}{A.~Gruzinov},
\newblock \bibinfo{title}{Photon-mass bound destroyed by vortices},
\newblock \bibinfo{journal}{Phys. Rev. Lett.} \bibinfo{volume}{98}
  (\bibinfo{year}{2007}) \bibinfo{pages}{010402}.
\bibitem[{Liu and Shao(2012)}]{liushao2012}
\bibinfo{author}{L.~X. Liu}, \bibinfo{author}{C.~G. Shao},
\newblock \bibinfo{title}{Re-estimatation of the upper limit on the photon mass
  with the solar wind method},
\newblock \bibinfo{journal}{Chin. Phys. Lett.} \bibinfo{volume}{29}
  (\bibinfo{year}{2012}) \bibinfo{pages}{111401}.
\bibitem[{Burlaga et~al.(1998)Burlaga, Ness, Wang, and
  Sheeley}]{burlaganesswangsheeley1998}
\bibinfo{author}{L.~F. Burlaga}, \bibinfo{author}{N.~F. Ness},
  \bibinfo{author}{Y.~M. Wang}, \bibinfo{author}{N.~Sheeley},
\newblock \bibinfo{title}{Heliospheric magnetic field strength out to 66 {AU}:
  {V}oyager 1, 1978–1996},
\newblock \bibinfo{journal}{J. Geophys. Res.} \bibinfo{volume}{103}
  (\bibinfo{year}{1998}) \bibinfo{pages}{23723}.
\bibitem[{Ness and Burlaga(2001)}]{nessburlaga2001}
\bibinfo{author}{N.~F. Ness}, \bibinfo{author}{L.~F. Burlaga},
\newblock \bibinfo{title}{Spacecraft studies of interplanetary magnetic field},
\newblock \bibinfo{journal}{J. Geophys. Res.} \bibinfo{volume}{106}
  (\bibinfo{year}{2001}) \bibinfo{pages}{15803}.
\bibitem[{Burlaga et~al.(2003)Burlaga, Ness, Wang, and
  Sheeley}]{burlaganesswangsheeley2003}
\bibinfo{author}{L.~F. Burlaga}, \bibinfo{author}{N.~F. Ness},
  \bibinfo{author}{Y.~M. Wang}, \bibinfo{author}{N.~R. Sheeley},
\newblock \bibinfo{title}{Voyager 1 studies of the {HMF} to 81 {AU} during the
  ascending phase of solar cycle 23},
\newblock in: \bibinfo{editor}{M.~Velli}, \bibinfo{editor}{R.~Bruno},
  \bibinfo{editor}{F.~Malara} (Eds.), \bibinfo{booktitle}{{P}roc.
  10\textsuperscript{th} Int. Solar Wind Conf.}, volume \bibinfo{volume}{679},
  \bibinfo{publisher}{American Institute of Physics},
  \bibinfo{address}{Melville}, \bibinfo{year}{2003}, p.~\bibinfo{pages}{39}.
  \bibinfo{note}{17-21 June 2002 Pisa}.
\bibitem[{Escoubet et~al.(1997)Escoubet, Schmidt, and Goldstein}]{esscgo97}
\bibinfo{author}{C.~P. Escoubet}, \bibinfo{author}{R.~Schmidt},
  \bibinfo{author}{M.~L. Goldstein},
\newblock \bibinfo{title}{{C}luster - science and mission overview},
\newblock \bibinfo{journal}{Space Sc. Rev.} \bibinfo{volume}{79}
  (\bibinfo{year}{1997}) \bibinfo{pages}{11}.
\bibitem[{Dunlop et~al.(2002)}]{dubaglro02breve}
\bibinfo{author}{M.~W. Dunlop}, et~al.,
\newblock \bibinfo{title}{Four-point {C}luster application of magnetic field
  analysis tools: The curlometer},
\newblock \bibinfo{journal}{J. Geophys. Res. (Space Physics)}
  \bibinfo{volume}{107} (\bibinfo{year}{2002}) \bibinfo{pages}{1384}.
\bibitem[{Robert et~al.(1998)}]{roduroch98breve}
\bibinfo{author}{P.~Robert}, et~al.,
\newblock \bibinfo{title}{Accuracy of current density determination},
\newblock in: \bibinfo{editor}{G.~Paschmann}, \bibinfo{editor}{P.~Daly} (Eds.),
  \bibinfo{booktitle}{Analysis method for multi-spacecraft data}, Int. Space
  Science Inst.{,} Bern, \bibinfo{publisher}{ESA},
  \bibinfo{address}{Noordwijk}, \bibinfo{year}{1998}, p. \bibinfo{pages}{395}.
\bibitem[{Balogh et~al.(1997)}]{baetal97breve}
\bibinfo{author}{A.~Balogh}, et~al.,
\newblock \bibinfo{title}{The {C}luster magnetic field investigation},
\newblock \bibinfo{journal}{Space Sc. Rev.} \bibinfo{volume}{79}
  (\bibinfo{year}{1997}) \bibinfo{pages}{65}.
\bibitem[{R{\`e}me et~al.(2001)R{\`e}me, Aoustin, Bosqued, Dandouras, Lavraud,
  Sauvaud, Barthe, Bouyssou, Camus, Coeur-Joly, Cros, Cuvilo, Ducay,
  Garbarowitz, Medale, Penou, Romefort, Rouzaud, Vallat, Alcayd{\'e}, Jacquey,
  Mazelle, D'Uston, M{\"o}bius, Kistler, Crocker, Granoff, Mouikis, Popecki,
  Vosbury, Klecker, Hovestadt, Kucharek, Kuenneth, Paschmann, Scholer, Sckopke,
  Carlson, Curtis, Ingraham, Lin, McFadden, Parks, Phan, Formisano, Amata,
  Bavassano-Cattaneo, Baldetti, Bruno, Chionchio, di~Lellis, Marcucci,
  Pallocchia, Korth, Daly, Graeve, Rosenbauer, Vasyliunas, McCarthy, Wilber,
  Eliasson, Lundin, Olsen, Shelley, Fuselier, Ghielmetti, Lennartsson,
  Escoubet, Balsiger, Friedel, Cao, Kovrazhkin, Papamastorakis, Pellat,
  Scudder, and Sonnerup}]{reetal01}
\bibinfo{author}{H.~R{\`e}me}, \bibinfo{author}{C.~Aoustin},
  \bibinfo{author}{J.~M. Bosqued}, \bibinfo{author}{I.~Dandouras},
  \bibinfo{author}{B.~Lavraud}, \bibinfo{author}{J.~A. Sauvaud},
  \bibinfo{author}{A.~Barthe}, \bibinfo{author}{J.~Bouyssou},
  \bibinfo{author}{T.~Camus}, \bibinfo{author}{O.~Coeur-Joly},
  \bibinfo{author}{A.~Cros}, \bibinfo{author}{J.~Cuvilo},
  \bibinfo{author}{F.~Ducay}, \bibinfo{author}{Y.~Garbarowitz},
  \bibinfo{author}{J.~L. Medale}, \bibinfo{author}{E.~Penou},
  \bibinfo{author}{H.~P.~D. Romefort}, \bibinfo{author}{J.~Rouzaud},
  \bibinfo{author}{C.~Vallat}, \bibinfo{author}{D.~Alcayd{\'e}},
  \bibinfo{author}{C.~Jacquey}, \bibinfo{author}{C.~Mazelle},
  \bibinfo{author}{C.~D'Uston}, \bibinfo{author}{E.~M{\"o}bius},
  \bibinfo{author}{L.~M. Kistler}, \bibinfo{author}{K.~Crocker},
  \bibinfo{author}{M.~Granoff}, \bibinfo{author}{C.~Mouikis},
  \bibinfo{author}{M.~Popecki}, \bibinfo{author}{M.~Vosbury},
  \bibinfo{author}{B.~Klecker}, \bibinfo{author}{D.~Hovestadt},
  \bibinfo{author}{H.~Kucharek}, \bibinfo{author}{E.~Kuenneth},
  \bibinfo{author}{G.~Paschmann}, \bibinfo{author}{M.~Scholer},
  \bibinfo{author}{N.~Sckopke}, \bibinfo{author}{E.~S. C.~W. Carlson},
  \bibinfo{author}{D.~W. Curtis}, \bibinfo{author}{C.~Ingraham},
  \bibinfo{author}{R.~P. Lin}, \bibinfo{author}{J.~P. McFadden},
  \bibinfo{author}{G.~K. Parks}, \bibinfo{author}{T.~Phan},
  \bibinfo{author}{V.~Formisano}, \bibinfo{author}{E.~Amata},
  \bibinfo{author}{M.~B. Bavassano-Cattaneo}, \bibinfo{author}{P.~Baldetti},
  \bibinfo{author}{R.~Bruno}, \bibinfo{author}{G.~Chionchio},
  \bibinfo{author}{A.~di~Lellis}, \bibinfo{author}{M.~F. Marcucci},
  \bibinfo{author}{G.~Pallocchia}, \bibinfo{author}{A.~Korth},
  \bibinfo{author}{P.~W. Daly}, \bibinfo{author}{B.~Graeve},
  \bibinfo{author}{H.~Rosenbauer}, \bibinfo{author}{V.~Vasyliunas},
  \bibinfo{author}{M.~McCarthy}, \bibinfo{author}{M.~Wilber},
  \bibinfo{author}{L.~Eliasson}, \bibinfo{author}{R.~Lundin},
  \bibinfo{author}{S.~Olsen}, \bibinfo{author}{E.~G. Shelley},
  \bibinfo{author}{S.~Fuselier}, \bibinfo{author}{A.~G. Ghielmetti},
  \bibinfo{author}{W.~Lennartsson}, \bibinfo{author}{C.~P. Escoubet},
  \bibinfo{author}{H.~Balsiger}, \bibinfo{author}{R.~Friedel},
  \bibinfo{author}{J.~B. Cao}, \bibinfo{author}{R.~A. Kovrazhkin},
  \bibinfo{author}{Papamastorakis}, \bibinfo{author}{R.~Pellat},
  \bibinfo{author}{J.~Scudder}, \bibinfo{author}{B.~Sonnerup},
\newblock \bibinfo{title}{First multispacecraft ion measurements in and near
  the {E}arth's magnetosphere with the identical {C}luster ion spectrometry
  ({CIS}) experiment},
\newblock \bibinfo{journal}{Ann. Geophys.} \bibinfo{volume}{19}
  (\bibinfo{year}{2001}) \bibinfo{pages}{1303}.
\bibitem[{Johnstone et~al.(1997)Johnstone, Alsop, Burge, Carter, Coates, Coker,
  Fazakerley, Grande, Gowen, Gurgiolo, Hancock, Nartheim, Preece, Sheather,
  Winningham, and Woodliffe}]{joetal97}
\bibinfo{author}{A.~D. Johnstone}, \bibinfo{author}{C.~Alsop},
  \bibinfo{author}{S.~Burge}, \bibinfo{author}{P.~J. Carter},
  \bibinfo{author}{A.~J. Coates}, \bibinfo{author}{A.~J. Coker},
  \bibinfo{author}{A.~N. Fazakerley}, \bibinfo{author}{M.~Grande},
  \bibinfo{author}{R.~A. Gowen}, \bibinfo{author}{C.~Gurgiolo},
  \bibinfo{author}{B.~K. Hancock}, \bibinfo{author}{B.~Nartheim},
  \bibinfo{author}{A.~Preece}, \bibinfo{author}{P.~H. Sheather},
  \bibinfo{author}{J.~D. Winningham}, \bibinfo{author}{R.~D. Woodliffe},
\newblock \bibinfo{title}{Peace: a plasma electron and current experiment},
\newblock \bibinfo{journal}{Space Sc. Rev.} \bibinfo{volume}{79}
  (\bibinfo{year}{1997}) \bibinfo{pages}{351}.
\bibitem[{Bieber et~al.(1987)Bieber, Evenson, and Matthaeus}]{bieberetal1987}
\bibinfo{author}{J.~W. Bieber}, \bibinfo{author}{P.~A. Evenson},
  \bibinfo{author}{W.~H. Matthaeus},
\newblock \bibinfo{title}{Magnetic helicity of the {P}arker field},
\newblock \bibinfo{journal}{Astrophys. J.} \bibinfo{volume}{315}
  (\bibinfo{year}{1987}) \bibinfo{pages}{700}.
\bibitem[{Paschmann et~al.(1998)Paschmann, Fazakerley, and Schwartz}]{pafasc98}
\bibinfo{author}{G.~Paschmann}, \bibinfo{author}{A.~N. Fazakerley},
  \bibinfo{author}{S.~J. Schwartz},
\newblock \bibinfo{title}{Moments of plasma velocity distribution},
\newblock in: \bibinfo{editor}{G.~Paschmann}, \bibinfo{editor}{P.~Daly} (Eds.),
  \bibinfo{booktitle}{Analysis method for multi-spacecraft data}, Int. Space
  Science Inst.{,} Bern, \bibinfo{publisher}{ESA},
  \bibinfo{address}{Noordwijk}, \bibinfo{year}{1998}, p. \bibinfo{pages}{125}.
\bibitem[{Taylor(1997)}]{taylor1997}
\bibinfo{author}{J.~R. Taylor}, \bibinfo{title}{An introduction to error
  analysis: the study of uncertainties in physical measurements},
  \bibinfo{publisher}{University Science Books}, \bibinfo{address}{Mill
  Valley}, \bibinfo{year}{1997}. \bibinfo{note}{2nd ed.}
\bibitem[{Ryutov(2009)}]{ry09}
\bibinfo{author}{D.~D. Ryutov},
\newblock \bibinfo{title}{Relating the {P}roca photon mass and cosmic vector
  potential via solar wind},
\newblock \bibinfo{journal}{Phys. Rev. Lett.} \bibinfo{volume}{103}
  (\bibinfo{year}{2009}) \bibinfo{pages}{201803}.
\bibitem[{Graham et~al.(2015)Graham, Khotyaintsev, Vaivads, Andre, Lindqvist,
  {Le Contel}, Ergun, Goodrich, Torbert, Russell, Magnes, Pollock, Mauk, and
  Fuselier}]{grahametal2015}
\bibinfo{author}{D.~B. Graham}, \bibinfo{author}{Y.~V. Khotyaintsev},
  \bibinfo{author}{A.~Vaivads}, \bibinfo{author}{M.~Andre},
  \bibinfo{author}{P.~Lindqvist}, \bibinfo{author}{O.~{Le Contel}},
  \bibinfo{author}{R.~E. Ergun}, \bibinfo{author}{K.~Goodrich},
  \bibinfo{author}{R.~B. Torbert}, \bibinfo{author}{C.~T. Russell},
  \bibinfo{author}{W.~Magnes}, \bibinfo{author}{C.~Pollock},
  \bibinfo{author}{B.~H. Mauk}, \bibinfo{author}{S.~A. Fuselier},
  \bibinfo{title}{{MMS} observations on the ion diffusion region on 30
  {O}ctober 2015}, \bibinfo{year}{2015}. \bibinfo{note}{{P}oster at AGU Fall
  Meeting, 14-18 December 2015 San Francisco}.

\end{thebibliography}

\end{document}